\documentclass[runningheads]{llncs}
\usepackage{multirow}
\usepackage{graphicx}
\usepackage{array}

\newcolumntype{P}[1]{>{\centering\arraybackslash}p{#1}}

\usepackage{amsmath}
\usepackage{xcolor}
\usepackage[normalem]{ulem}

\begin{document}
\title{Modeling Market States with Clustering and State Machines}
\titlerunning{Modeling Market States with Clustering and State Machines}
%
\author{Christian Oliva \and Silviu Gabriel Tinjala}
\authorrunning{Oliva, C. and Tinjala, S.G.}

%
%

\institute{Grupo de Neurocomputación Biológica, Departamento de Ingeniería Informática, Escuela Politécnica Superior, Universidad Autónoma de Madrid, Spain
\email{christian.oliva@uam.es,}
\email{silviu.tinjala@estudiante.uam.es}}

\maketitle              
\begin{abstract}

This work introduces a new framework for modeling financial markets through an interpretable probabilistic state machine. By clustering historical returns based on momentum and risk features across multiple time horizons, we identify distinct market states that capture underlying regimes, such as expansion phase, contraction, crisis, or recovery. From a transition matrix representing the dynamics between these states, we construct a probabilistic state machine that models the temporal evolution of the market. This state machine enables the generation of a custom distribution of returns based on a mixture of Gaussian components weighted by state frequencies. We show that the proposed benchmark significantly outperforms the traditional approach in capturing key statistical properties of asset returns, including skewness and kurtosis, and our experiments across random assets and time periods confirm its robustness.

\keywords{Financial modeling, Market regimes, Momentum clustering}
\end{abstract}
\section{Introduction}
\label{sec:introduction}

Financial markets rule the global economy, serving as platforms where capital is allocated, and prices reflect the collective sentiment and information of countless participants. These markets encompass equities, funds, commodities, etc., and are complex, adaptive systems influenced by a multitude of economic, political, and psychological factors.

Forecasting market behavior remains one of the most challenging goals in finance. Despite the wealth of historical data and advances in economic techniques, accurately predicting market movements is notoriously difficult due to the complex dynamics (and often chaotic nature) of the market. There is always a need to develop new strategies in response to new information. One conceptual framework that has gained attention for understanding the cyclical patterns in market behavior is the \textit{market clock}. This model visualizes the market as moving through a cyclical series of distinct states, such as expansion, peak, contraction, and trough, analogous to the hours on a clock. These states correspond to varying investor sentiments and economic conditions; thus, the \textit{market clock} serves as a heuristic tool to anticipate these transitions between market regimes.

Building on this concept, recent advances in AI offer promising avenues for identifying and predicting market states. Machine Learning (ML) models, capable of uncovering complex patterns in large amounts of data, can also be used to identify market conditions and forecast transitions between regimes. In this work, we present a benchmark for modeling market states in a simple, interpretable, and robust manner, using basic clustering algorithms and probabilistic state machines. We show that asset behaviors can be captured with this approach, revealing structural patterns that correspond to different phases of the market cycle. Our aim with this benchmark is not to propose a new investment strategy or optimize portfolios, but to bridge the gap between purely data-driven techniques and economically meaningful representations of market regimes, aiming for a better understanding and more informed decision-making.

The remaining of the article is organized as follows. In Section \ref{sec:related_work}, we briefly review the state of the art in clustering and forecasting market states. Then, in Section \ref{sec:methodology}, we present the benchmark for modeling market states, detailing the use of clustering to create probabilistic state machines, and describe and analyze the incoming results. Finally, Section \ref{sec:conclusion} presents our conclusions and discusses potential directions for further research.

\section{Related work}
\label{sec:related_work}

Many time series models presented in the literature tried to capture the changes in market regimes. Among the classical methods, it is worth mentioning the Markov Switching models \cite{Markov_chain_89}, which model the output of an autoregression as the outcome of a discrete-state Markov process. However, these models are highly susceptible to the curse of dimensionality; therefore, it is common to rely on variational inference techniques \cite{Kim_99,Tsay_2005}. In a multivariate context, different market states are affected not only by gains and losses, but also by their internal dynamics. Classical approaches assume a stationary correlation structure \cite{Black01091992}, but it is well established that correlations among stocks are not constant over time and tend to increase in periods of high market volatility \cite{ang_2015,Cizeau01022001,lin_94}. In addition, momentum is one of the most studied phenomena in financial markets, referring to the tendency of assets with strong past performance to continue the trend well in the near future, and vice versa \cite{jegadeesh_93}. 

In recent years, researchers have explored the use of momentum-based clustering techniques to create trading and forecasting strategies \cite{baltas2013momentum,peachavanish2016stock,WU2022109358}. In addition, clustering has also gained popularity as an alternative to identify market regimes. The most common strategies are based on finding the number of volatility regimes in non-stationary financial time series \cite{Procacci_2019}, examining relationships in expected log-returns over time \cite{Prakash_2021}, or analyzing changes in the correlation structure \cite{BASALTO2005196,Hendricks01112016,Munnix_2012,pharasi2020marketstatesnewunderstanding}. However, it is uncommon to find research that explicitly combines momentum and clustering in the context of modeling market states. 

\section{Modeling market states}
\label{sec:methodology}

In this work, we present a new benchmark for modeling market states with clustering and state machines. In Section \ref{subsec:clustering}, we describe the attribute selection process for clustering, aiming to distinguish between different asset behaviors. Then, in Section \ref{subsec:building}, we show how the generated clusters are used to build the probabilistic state machine and its interpretability. Finally, in Section \ref{subsec:experiment}, we evaluate the robustness of the proposed benchmark in different scenarios.

\subsection{Clustering}
\label{subsec:clustering}

Clustering is designed to find patterns in unsupervised scenarios. In our case, the goal is to separate asset returns into different clusters with common behavior (i.e. similar trends and risk), as financial markets are known to move in cyclical patterns often described by the \textit{market clock}. This reflects recurring phases such as recovery, expansion, slowdown, and contraction, each characterized by distinct statistical properties in asset returns. For example, the expansion phase is typically associated with positive expected returns and moderate to low volatility. Thus, attribute selection should incorporate both momentum and risk characteristics.

\subsubsection*{Momentum.}

In our experiments, we use the momentum on different days to evaluate the mentioned tendency:

\begin{equation}
    Mom_t = \frac{P_T}{P_{T-t}}, \label{eq:momentum}
\end{equation}

\noindent where $P_T$ is the actual price at time $T$ and $P_{T-t}$ is the price $t$ days before. However, instead of using raw momentum, we use the logarithmic version:

\begin{equation}
    \log(Mom_t) = \log(P_T) - \log(P_{T-t}) = \sum_{i=t}^T r_i,
\end{equation}

\noindent where $r_i$ is the logarithmic return of the asset on day $i$. In our experiments, we compute log-momentum at different $t$, specifically 5, 10, 20, 30, 40, and 50 days. 

\subsubsection*{Risk.} 
\begin{figure}[!bht]
    \centering
    \includegraphics[width=0.32\linewidth]{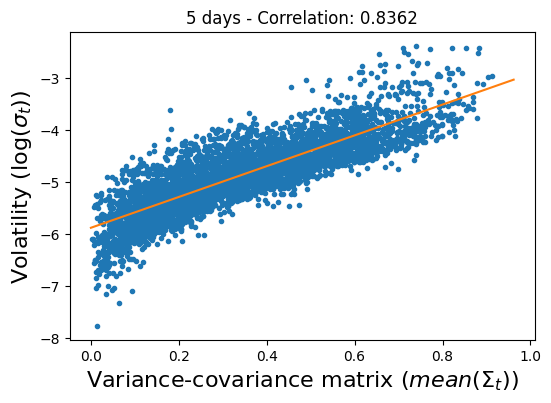}
    \includegraphics[width=0.32\linewidth]{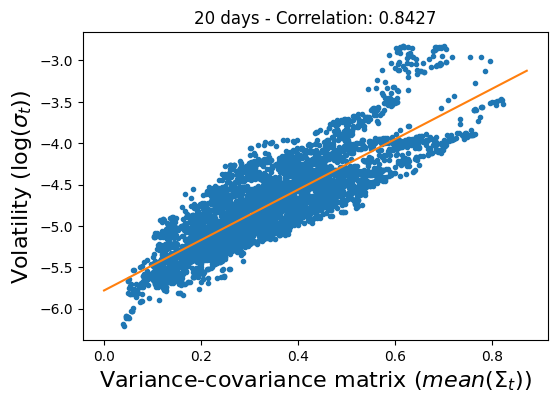}
    \includegraphics[width=0.32\linewidth]{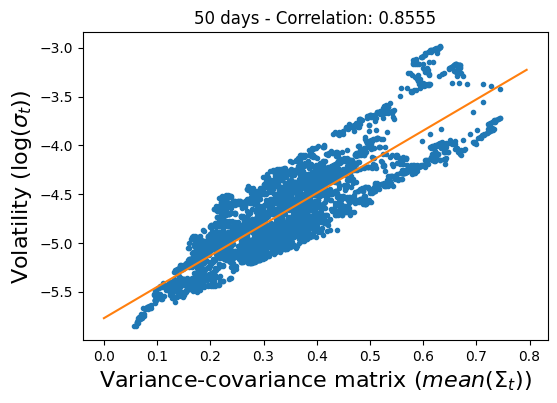}
    \caption{Log-volatility ($\log(\sigma_t)$) versus the mean of the variance-covariance matrix ($mean(\Sigma_t)$) at different rolling windows. From left to right, $t=5$, $t=20$, and $t=50$. Note that, in the three scenarios, the correlation between $\log(\sigma_t)$ and $mean(\Sigma_t)$ is higher than 0.83.}
    \label{fig:Sigma_vs_sigma}
\end{figure}

In financial economics, the variance-covariance matrix of the returns of assets ($\Sigma$) is the main tool for quantifying the market risk. It captures both individual variances and how assets move together over time (covariances). When evaluated in a rolling window, its average value $mean(\Sigma_t)$ offers a direct risk measure. However, directly estimating $\Sigma_t$ in high-dimensional settings can be computationally intensive; therefore, we propose using the standard deviation of a market index (i.e. S\&P500) as a proxy for market risk. This idea is empirically reinforced with the evaluation of different rolling windows at $t$ days over $mean(\Sigma_t)$ and $\sigma_t$, as shown in Figure \ref{fig:Sigma_vs_sigma}.

As shown in the figures, $\log(\sigma_t)$ is highly and positively correlated with $mean(\Sigma_t)$, suggesting that index volatility effectively encapsulates the average behavior of risk at the underlying asset level. In addition, it has much lower computational load; thus, for evaluating the risk, we use the standard deviation:

\begin{equation}
    \mu_t = \frac{1}{T} \sum_{i=1}^T r_i,
\end{equation}

\begin{equation}
    Risk_t = \sqrt{\frac{1}{T} \sum_{i=t}^T (r_i - \mu_t)^2},
\end{equation}

\noindent where $\mu_t$ is the expected return in the window [T-t, T]. Again, we use risk at different $t$, specifically 5, 10, 20, 30, 40, and 50 days. Table \ref{tab:ejemplo_x_train} shows the first two input data with the selected attributes of momentum and risk.

\begin{table}[!bht]
    \centering
    \caption{First two input data from the selected attributes for clustering}
    \begin{tabular}{rccccccc}
        \textbf{Date} & \textbf{$Mom_5$} & \textbf{$Risk_5$} & \textbf{$Mom_{10}$} & \textbf{$Risk_{10}$} & $\cdot \cdot \cdot$ & \textbf{$Mom_{50}$} & \textbf{$Risk_{50}$} \\
        \hline
        \textbf{2007-03-16} & -0.012 & 0.010 & 0.003 & 0.010 & \multirow{2}{*}{$\cdot \cdot\cdot$} & -0.016 & 0.008 \\
        \textbf{2007-03-19} & -0.002 & 0.012 & 0.024 & 0.010 & & -0.006 & 0.009 \\
        \hline
    \end{tabular}
    \label{tab:ejemplo_x_train}
\end{table}

\subsubsection*{K-Means.} 
\begin{figure}
    \centering
    \includegraphics[width=\linewidth]{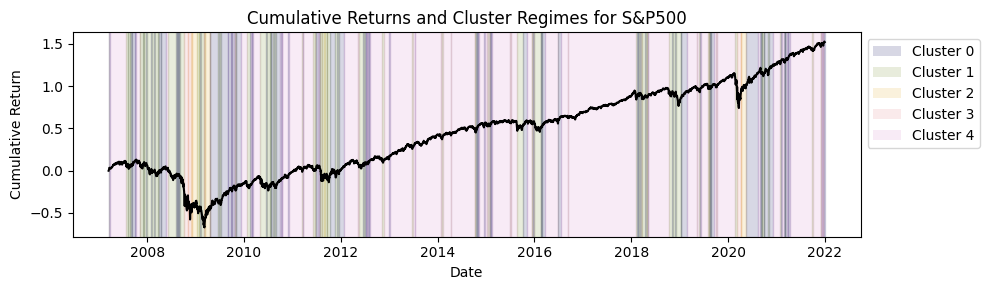}
    \caption{Cumulative returns and cluster regimes for S\&P500 index from 2007 to 2022 (training data).}
    \label{fig:ejemplo_clustering_sp500}
\end{figure}

K-Means is a widely used clustering algorithm that splits the data into a predefined number of clusters $K$. In our case, each asset on each day is represented as a feature vector $\mathbf{x}$ composed of multiple log-momentum and risk values at different times. K-Means will assign each vector $\mathbf{x}$ to the nearest cluster centroid in this multidimensional space. This allows the identification of distinct asset behaviors, each representing a specific combination of momentum and risk characteristics. Thus, each day in the training data (see Fig. \ref{tab:ejemplo_x_train}) is assigned to one of the $K$ clusters based on its proximity to the nearest centroid. Figure \ref{fig:ejemplo_clustering_sp500} shows the clustering of each day applied to the S\&P500 index from 2007 to 2022 (training data) with $K=5$. Note that all clustering operations are always performed on standardized data.

Once all days in the training data are assigned to a cluster, we can interpret each cluster by analyzing the centroids. Table \ref{tab:clusters_interpretability} shows them to proceed with its interpretation.

\begin{table}[!bht]
    \centering
    \caption{Cluster centroids of the example shown in Figure \ref{fig:ejemplo_clustering_sp500}}
    \begin{tabular}{ccccccccccccc}
         \textbf{Cluster} & $Mom_5$ & $Mom_{10}$ & $\cdot \cdot \cdot$ & $Mom_{40}$ & $Mom_{50}$ & $Risk_5$ & $Risk_{10}$ & $\cdot \cdot \cdot$ & $Risk_{40}$ & $Risk_{50}$ \\
         \hline
         \textbf{0} & 0.018 & 0.026 & \multirow{5}{*}{$\cdot \cdot \cdot$} & 0.040 & 0.050 & 0.010 & 0.012 & \multirow{5}{*}{$\cdot \cdot \cdot$} & 0.013 & 0.014 \\
         \textbf{1} & -0.018 & -0.029 & & -0.046 & -0.047 & 0.015 & 0.015 & & 0.013 & 0.013 \\
         \textbf{2} & 0.027 & 0.043 & & -0.036 & -0.093 & 0.021 & 0.025 & & 0.035 & 0.036 \\
         \textbf{3} & -0.043 & -0.087 & & -0.277 & -0.289 & 0.051 & 0.050 & & 0.040 & 0.036 \\
         \textbf{4} & 0.004 & 0.009 & & 0.038 & 0.046 & 0.006 & 0.006 & & 0.007 & 0.007 \\
         \hline
    \end{tabular}
    \label{tab:clusters_interpretability}
\end{table}

\begin{figure}
    \centering
    \includegraphics[width=\linewidth]{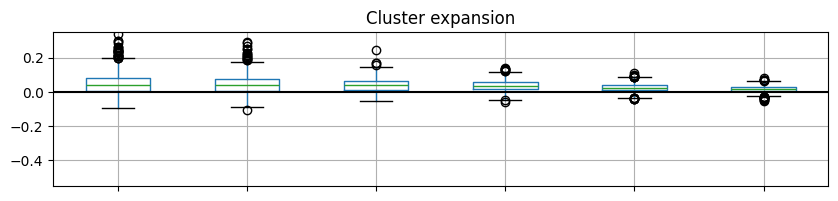}
    \includegraphics[width=\linewidth]{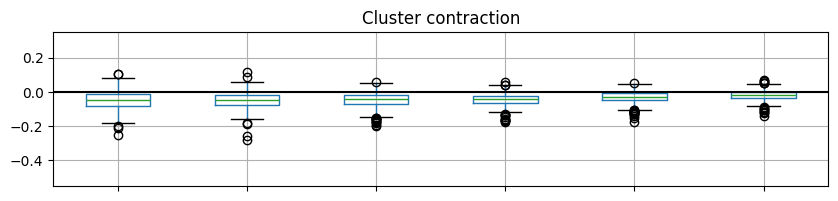}
    \includegraphics[width=\linewidth]{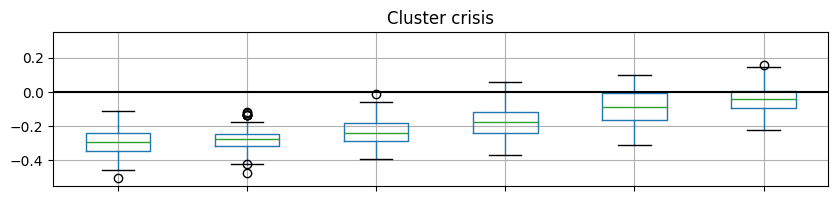}
    \includegraphics[width=\linewidth]{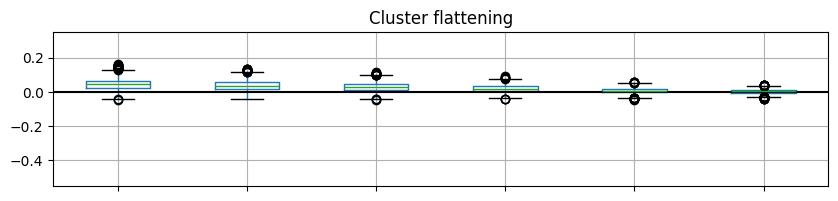}
    \includegraphics[width=\linewidth]{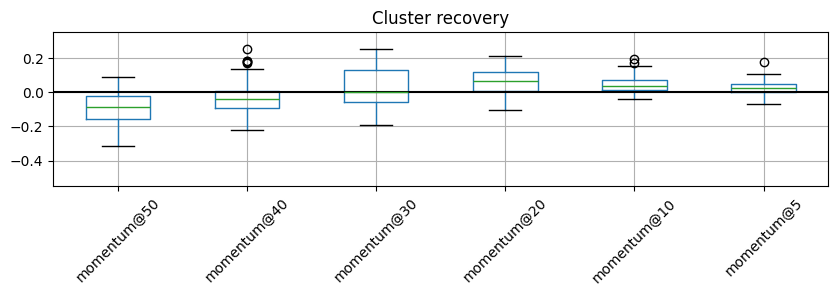}
    \caption{Box plot of momentum indicators of different states identified by the clustering. Each plot corresponds to a specific cluster: from top to bottom, \textit{expansion}, \textit{contraction}, \textit{crisis}, \textit{flattening}, and \textit{recovery}.}
    \label{fig:boxplot_clusters}
\end{figure}

Each cluster can be interpreted as follows: cluster 0 has high positive momentum and decreasing risk from early to late days. This suggests sustained upward movement, signaling market stabilization. Thus, we could call this cluster \textit{``expansion''}. Cluster 1 has the opposite effect, it has negative momentum and increasing risk from early to late days. This is like an initial phase of contraction, where volatility spikes briefly (\textit{``contraction''}). Cluster 2 represents a rebound, with momentum moving from negative to positive in short terms and with risk being strongly high (\textit{``recovery''}). Cluster 3 is a deep contraction, where the momentum is strongly negative at all times and the risk is extremely high (\textit{``crisis''}). Finally, cluster 4 seems to represent a peak, where momentum tends to 0 when is later. The risk has very low values, indicating a possible change in tendency (\textit{``flattening''}). Figure \ref{fig:boxplot_clusters} represents the box plot of momentum indicators of the different clusters.

\subsection{Building the State Machine}
\label{subsec:building}

\begin{figure}
    \centering
    \includegraphics[width=0.8\linewidth]{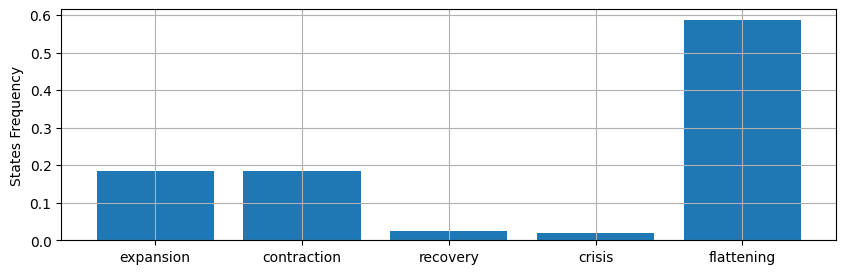}
    \caption{Probability of being in each state}
    \label{fig:states_frequency}
\end{figure}

Once all days in the training data are assigned to a cluster (see Fig. \ref{fig:ejemplo_clustering_sp500}), and interpreting these clusters, we construct a transition matrix that models the dynamics between clusters over time. Specifically, we track transitions from the cluster $c$ at time $t$ to the following cluster $c'$ at time $t+1$. This results in a square matrix $M$ of size $K\times K$, in which $M_{ij}$ counts how many times there is a transition from the $i$-th cluster to the $j$-th cluster. From this transition matrix, we extract two key elements. First, counting the frequency of the clusters (\textit{freq}) estimates the marginal probability of being in a given state, as shown in Figure \ref{fig:states_frequency}.

\begin{figure}
    \centering
    \includegraphics[width=0.8\linewidth]{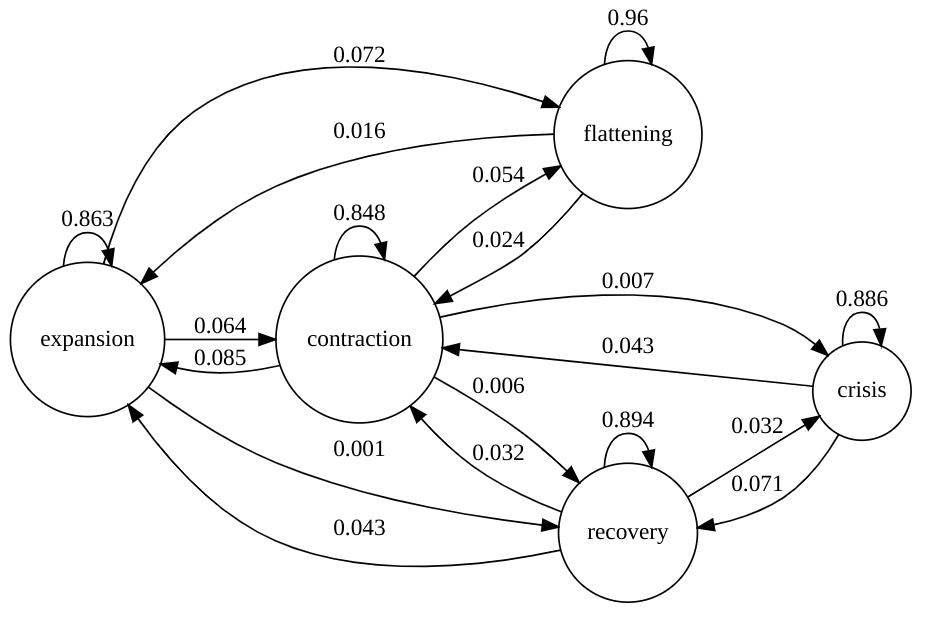}
    \caption{Probabilistic State Machine extracted from the transition matrix}
    \label{fig:automata}
\end{figure}

As shown in the figure, the probability of being in a \textit{flattening} state dominates with almost 60\% probability, while being in a \textit{crisis} state remains below 2\%. Second, by normalizing each row of the transition matrix, we obtain the conditional probabilities of moving from the cluster $c$ to any other cluster $c'$. This defines the probabilistic rules governing state transitions and forms the state machine, as shown in Figure \ref{fig:automata}.

Each state can be interpreted by the statistical characteristics of its corresponding centroid, allowing the interpretation of transition probabilities. For example, if we focus on the \textit{expansion} state, it remains unchanged with 86.3\% probability, transits to \textit{contraction} with 6.4\%, transits to \textit{flattening} with 7.2\%, transits to \textit{recovery} with 0.1\% and there is no chance to transit directly to the \textit{crisis} state. The same interpretation can be made for the other states.

\begin{figure}
    \centering
    \includegraphics[width=\linewidth]{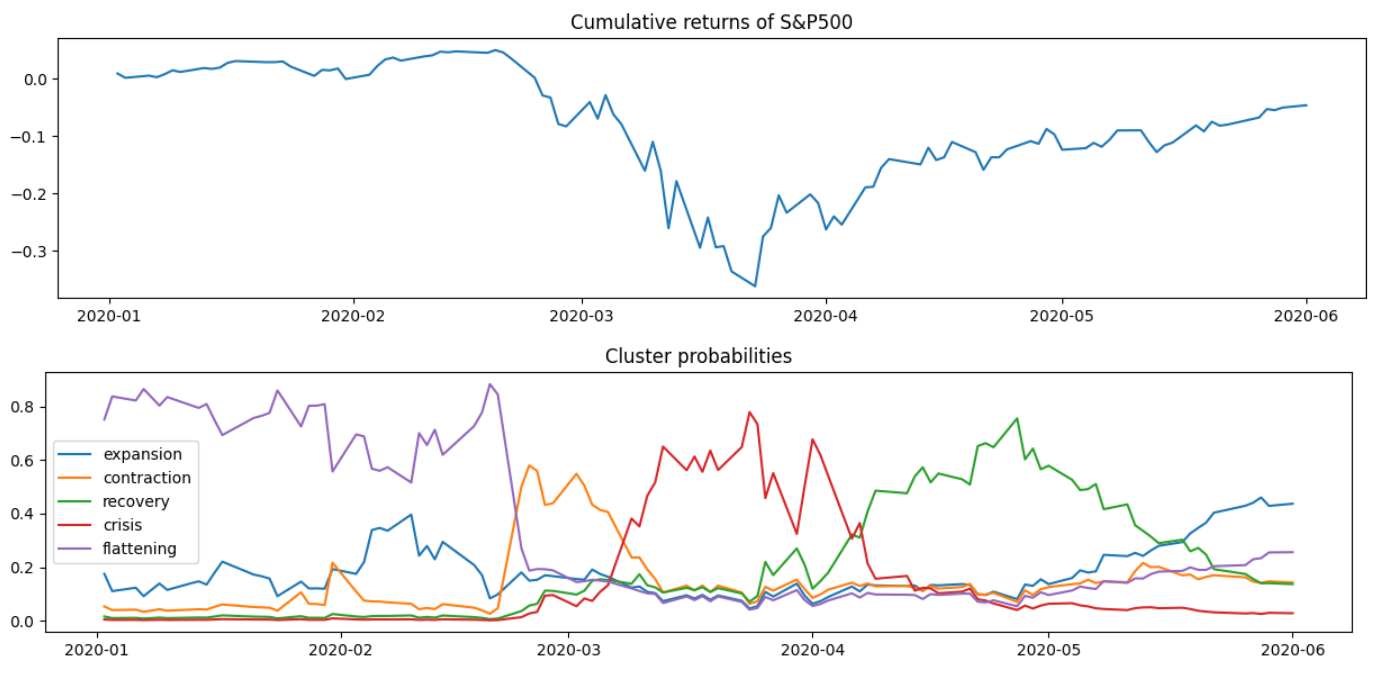}
    \caption{Top figure: Cumulative returns of S\&P500 index during the COVID-19 crisis. Bottom figure: Cluster probabilities in the same window.}
    \label{fig:cluster_probs}
\end{figure}

In addition to the state machine, represented in Figure \ref{fig:automata}, and the visualization of the cluster regimes (represented in Figure \ref{fig:ejemplo_clustering_sp500}), we can estimate the probability of each day $\mathbf{p}_t$ of being in each cluster as the inverse of the distance $\mathbf{d}_t$ from the centroids ($\mathbf{p}_t = 1 / \mathbf{d}_t$). Thus, we can also analyze the behavior of the probabilities over time. In Figure \ref{fig:cluster_probs}, we show an example of these probabilities $\mathbf{p}$ from January 2020 to June 2020, coinciding with the COVID-19 crisis.

As shown in the figures, during the previous months of the COVID crisis, the index state resides in the \textit{flattening} state (purple curve). Then, near March, the highest state probability switches to \textit{contraction} (orange curve) and in a few days it switches again to \textit{crisis} (red curve), as expected. Then, in April, the highest probability changes to \textit{recovery} (green curve), followed by a transition to \textit{expansion} (blue curve) near June. These shifts closely correspond to actual market events during the pandemic, validating the interpretability of the states.

In addition to the state machine and its probabilities, we can generate the overall distribution of the returns. We model the full return distribution as a mixture of Gaussian distributions, where each component corresponds to one of the clusters (i.e., each cluster has its own mean and variance). The key is that each cluster is weighted by its frequency in the training data (see Figure \ref{fig:states_frequency}), thus each Gaussian contributes according to that frequency.

\begin{equation}
    R_i \sim \sum_{i=1}^K c_i \ \mathcal{N}(\mu_i, \sigma_i),
\end{equation}

\noindent where $c_i$ is the $i$-th hot encoding component of $\mathbf{c}$, which has been generated from a categorical choice with probabilities \textit{freq} (see Fig. \ref{fig:states_frequency}). To analyze this proposal, Table \ref{tab:mean_std_skew_kurtosis} reports the first four moments (mean, standard deviation, skewness, and kurtosis) for the real distribution of the test data (\textit{Test S\&P500}), a normal distribution $\mathcal{N}(\mu, \sigma)$ (where $\mu$ and $\sigma$ are extracted from the complete training data) and our state machine distribution.

\begin{table}
\begin{minipage}[c]{0.42\linewidth}
    \centering
    \includegraphics[width=\linewidth]{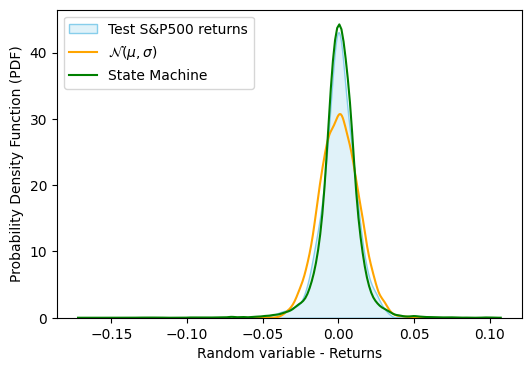}
    \label{fig:comparacion_distribuciones}
\end{minipage}%
\begin{minipage}[c]{0.58\linewidth}
    \centering
    \begin{tabular}{rccc}
         & \textit{Test S\&P500} & $\mathcal{N}(\mu, \sigma)$ & \textit{State Machine} \\
         \hline
         $\mu$ & 0.00039 & 0.00040 & 0.00037 \\
         $\sigma$ & 0.01251 & 0.01276 & 0.01295 \\
         \textbf{Skew.} & -0.3706 & 0.0204 & -1.4788 \\
         \textbf{Kurt.} & 13.527 & -0.069 & 19.6835 \\
        \hline
    \end{tabular}
    \label{tab:mean_std_skew_kurtosis}
\end{minipage}
\caption{Figure represents the normal distribution $\mathcal{N}(\mu, \sigma)$, and the distributions of the test data and the state machine. The table shows the values of the first four moments (mean, standard deviation, skewness, and kurtosis) for each distribution.}
\label{tab:mean_std_skew_kurtosis}
\end{table}

The results shown in the table confirm that, although both models preserve the mean and standard deviation relatively well, the state machine model substantially improves the modeling of higher-order moments. Specifically, it captures the negative skewness and excess kurtosis typical of financial returns, which are missed by the normal distribution. To provide a more rigorous comparison, Table \ref{tab:rigorous_comparison} reports three distribution distance metrics (Kullback-Leibler divergence, Kolmogorov-Smirnov statistic and Wasserstein distance) between the real test returns and each of the two distributions. Note that lower values indicate better fit.

\begin{table}[!bht]
    \centering
    \caption{Distributional distance metrics between the test real returns and each of the two models. Closer to 0.0 is better in all distances.}
    \begin{tabular}{rccc}
         \textbf{Distribution} & \textbf{Kolmogorov-Smirnov} & \textbf{Kullback-Leibler} & \textbf{Wasserstein} \\
         \hline
        \textit{$\mathcal{N}(\mu, \sigma)$} & 0.086 & 59.386 & 0.0021 \\
        \textit{State Machine} & 0.034 & 5.442 & 0.0009 \\
        \hline
    \end{tabular}
    \label{tab:rigorous_comparison}
\end{table}

These results highlight the better performance of the state machine in all metrics. It consistently achieves lower distance values, indicating that it better approximates the return distribution and better captures the complex dynamics of asset returns than the normal distribution $\mathcal{N}(\mu, \sigma)$. Having presented the state machine, in the following section (Sec. \ref{subsec:experiment}) we perform an additional experiment to evaluate this benchmark in different scenarios.

\subsection{Asset Modeling in different scenarios}
\label{subsec:experiment}

Once we have analyzed the performance of the benchmark in a single scenario with the S\&P500 index, we now extend our evaluation to a broader set of conditions to test its robustness. The goal of this experiment is to assess how the proposed state machine performs when applied to different assets and time periods, and whether its consistent advantage over the normal distribution ($\mathcal{N}(\mu, \sigma)$) persists. 

\begin{figure}
    \centering
    \includegraphics[width=\linewidth]{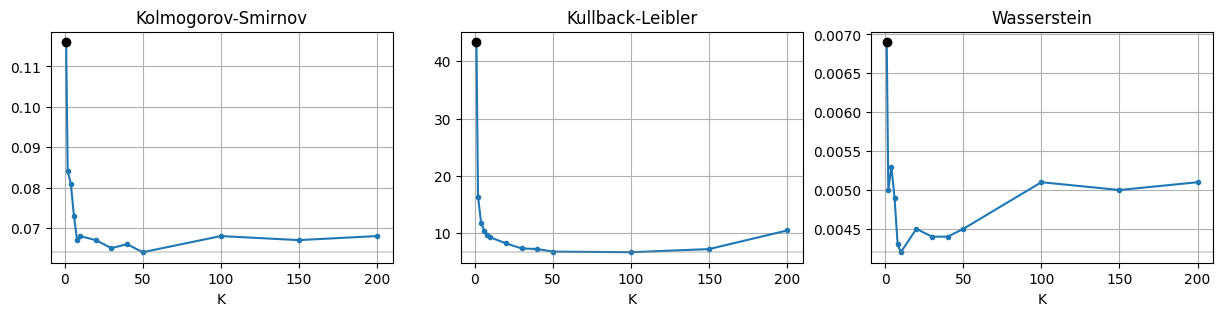}
    \caption{From left to right, the figures show the Kolmogorov-Smirnov, Kullback-Leibler, and Wasserstein distances for different $K$ values. The black dot represents the normal distribution $\mathcal{N}(\mu, \sigma)$.}
    \label{fig:summary_exp1}
\end{figure}

Thus, we implement a randomized testing framework. In each iteration, we randomly select an asset from the available dataset, as well as a random training period defined by a start and end date (the test set is fixed as the remaining data from the end of the training period up to the most recent available date). For each scenario, we train the full model, including clustering, the construction of the state machine, and the generation of the return distribution, and then compute the three distribution distance metrics (KL, Kolmogorov, and Wasserstein) to compare the generated distribution to that of the real test returns. To examine the effect of the number of clusters $K$ on the performance of our proposal, we repeat this experiment 500 times for each value of $K$ from 2 to 200. Figure \ref{fig:summary_exp1} shows a summary of the results. From left to right, the figures show the Kolmogorov-Smirnov, Kullback-Leibler, and Wasserstein distances. Note that the black dot in the three figures represents the normal distribution $\mathcal{N}(\mu, \sigma)$.

From the figures, we can observe, on the one hand, that the normal distribution $\mathcal{N}(\mu, \sigma)$ performs significantly worse than any of the state machines, even with the minimal number of clusters ($K=2$). Thus, the proposed approach consistently achieves substantially lower divergence and distance values. Regarding the choice of $K$, the performance of the state machine improves consistently as $K$ increases, with the most notable gains up to around $K=10$. Beyond that value, the improvements become increasingly marginal and, from approximately $K=50$ onward, they begin to deteriorate. 

\begin{figure}
    \centering
    \includegraphics[width=0.32\linewidth]{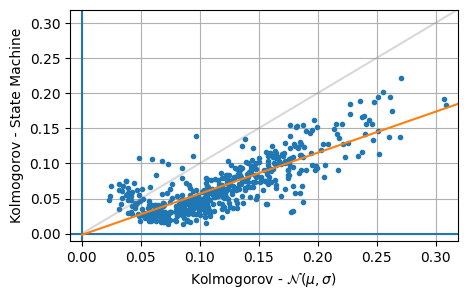}
    \includegraphics[width=0.32\linewidth]{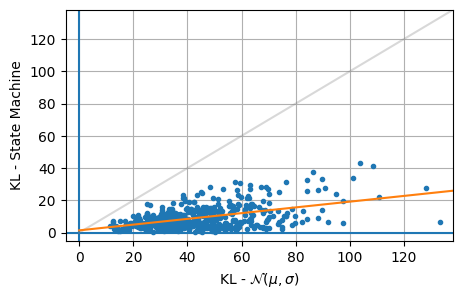}
    \includegraphics[width=0.32\linewidth]{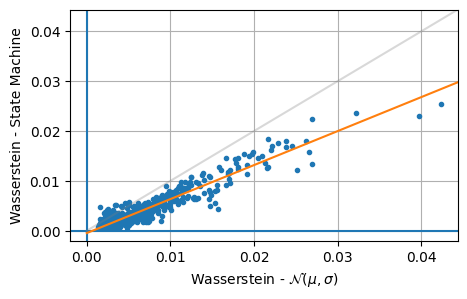}
    \caption{State machine distances versus normal distribution distances to the real test returns. From left to right, we show the results from Kolmogorov, KL, and Wasserstein distances. Note that the majority of the 500 random scenarios are below the diagonal, which indicates that the state machine yields lower distances than the normal distribution. In addition, orange line represents the linear regression of the two results. }
    \label{fig:normal_vs_state_machine}
\end{figure}

This highlights a discussion on the best choice for the number of clusters $K$. There is an important tradeoff between the state machine complexity and practical utility. Although larger $K$ values allow for more detailed modeling, they also reduce its interpretability. A moderate value, such as $K=10$, emerges as a robust and balanced choice. In fact, the Wasserstein distance identifies $K=10$ as the best value. To further validate the robustness of the proposed model, we show in Figure \ref{fig:normal_vs_state_machine} the performance of the state machine with $K=10$ relative to the normal distribution under the same scenarios. 

The results show that almost all points lie below the diagonal, indicating that the state machine consistently has lower distances than the normal distribution for the same scenarios. This pattern holds across the three distances, confirming that the improvement is not specific to a single evaluation criterion. Furthermore, we overlay a linear regression (orange line) to capture the general trend. As distances increase (perhaps due to more complex or harder return distributions), the gap between the two models widens in favor of the state machine.

\section{Conclusion}
\label{sec:conclusion}

In this work, we have presented a methodology for modeling financial asset behavior through an interpretable probabilistic framework. The main idea consists of clustering historical asset returns based on momentum and risk attributes across multiple time horizons, allowing the identification of market states. These clusters are then used to build a probabilistic state machine, which captures the dynamics of the market transitions and enables the generation of a custom return distribution based on a mixture of Gaussian components, one per state. We have shown that this model offers a more flexible and realistic alternative to the traditional normal distribution approach. 

Several key contributions emerge from this proposal. First, the interpretability of the state machine allows practitioners to qualitatively understand different phases of the market and their transition dynamics. Second, the probabilistic structure of state transitions opens the door to signal generation and strategy development, since the frequencies and probabilities of the clusters at each day can be monitored and used as new indicators. Third, the state machine demonstrates strong robustness. Throughout the experiment performed, which involves different assets, time periods, and values of $K$, the proposed state machine consistently outperforms the standard normal model in approximating the empirical return distribution.

Looking ahead, there are several promising future research areas. One direction involves extending the temporal range of the momentum and risk features beyond 50 days or incorporating additional attributes, such as volume, macroeconomic indicators, technical factors, etc. Furthermore, scaling this approach to a multi-asset setting may enable portfolio optimization, such as regime-aware asset allocation or risk management. In summary, the proposed state machine framework offers a flexible, interpretable, and robust tool for modeling financial markets, with a wide range of possible applications.

\section*{Acknowledgements}

This research was supported by grant PID2023-149669NB-I00 (MCIN/AEI and ERDF - ``A way of making Europe'').

%
%
%
\bibliographystyle{splncs04}
\bibliography{mybib}

\end{document}